\documentclass[twocolumn]{revtex4}
\usepackage{graphicx}

\begin{document}

\title{Hot electrons in a tunnel structure based on metal nanoclusters}

\author {V. V. Pogosov\footnote {Corresponding author:
vpogosov@zntu.edu.ua}$\dag$$\ddag$, E. V. Vasyutin$\dag$, A. V.
Babich$\dag$}

\address{$\dag$Department of Microelectronics and Semiconductor Devices,
Zaporozhye National Technical University, Zhukovsky Street 64,
Zaporozhye 69063, Ukraine \\ $\ddag$ Nanomaterials Laboratory,
Zaporozhye National University, Zhukovsky Street 66, Zaporozhye
69063, Ukraine}

\date{\today}

\begin{abstract}
We study the effect of temperature on the tunnel current in a
structure based on gold clusters taking  into consideration their
discrete electronic spectra. We suggest that an overheating of
electron subsystem leads to the disappearance of a current gap and
gradual smoothing of current--voltage curves that is observed
experimentally.
\end {abstract}



\maketitle

The nanodispersed metallic systems are prospective object of
nanotechnology \cite{Wang,Otero,Ohgi-2004}. Transport of electrical
charge across a nanoscale tunnel junction is accompanied by many
effects, such as the Coulomb blockade of the average current,
transfer of energy between electrons and ions, and consequent
heating of the junction. In nanometer scale devices, electron
transport can occur through well-resolved quantum states. If the
temperature is increased, the Coulomb and quantum staircases of
current are gradually smeared out by thermal fluctuations.

Simple tunnel construction can be schematically represented by the
distinctive ``sandwich'' \cite{Wang,Ohgi-2004}. It consists of a
thick gold film (emitter) covered by a dielectric one (with
dielectric constant $\epsilon \approx 3$).  Disc-shaped \cite{Wang}
or spherical-like \cite{Ohgi-2004} gold clusters are self-organized
on the dielectric layer. Also, a tip of STM is used in the capacity
of the third electrode (collector).

Some of the experimental features of the $I-V$ curves were
investigated in Ref. \cite{Nano-2006}, however, the fact of
smoothing of staircases for granule-molecule at low temperatures is
still not understood. Such a smoothing is typical for molecular
transistors \cite{1410}. Moreover, the observed current gap
decreases significantly as temperature increases from 5 K to 300 K
in structure based on disk-shaped cluster (Figure 2 in \cite{Wang}
for disk $2R \approx 4$\,nm).

In this letter, the temperature features of the $I-V$ curves are
explained by overheating of electron subsystem.

The number of atoms $N_{0}\simeq \{14,\, 10^{3}\}$ and
$\{100,\,600\}$ correspond to gold discs of monatomic thickness and
spheres whose radii vary in the range $2R \simeq \{1,\, 8.5\}$ and
$\{1.4,\, 2.8\}$\,nm, respectively. For given cluster sizes, the
condition $L \gg R$ is fulfilled, where $L$ is the free path length
for the electrons in the bulk of a metal.

The calculation of the electron spectrum in the cylindrical and
spherical wells of the mentioned sizes with finite deepness yields
different values for the spectrum discreteness in magic clusters
$\Delta\varepsilon_{p} =\varepsilon^{\rm LU} - \varepsilon^{\rm HO}$
(see Figure 1). In the nonmagic clusters the levels of lowest
unoccupied states coincide with those of highest occupied ones,
$\varepsilon^{\rm LU} = \varepsilon^{\rm HO}$ at $T=0$.
\begin{figure}
\centering
\includegraphics [width =.49\textwidth] {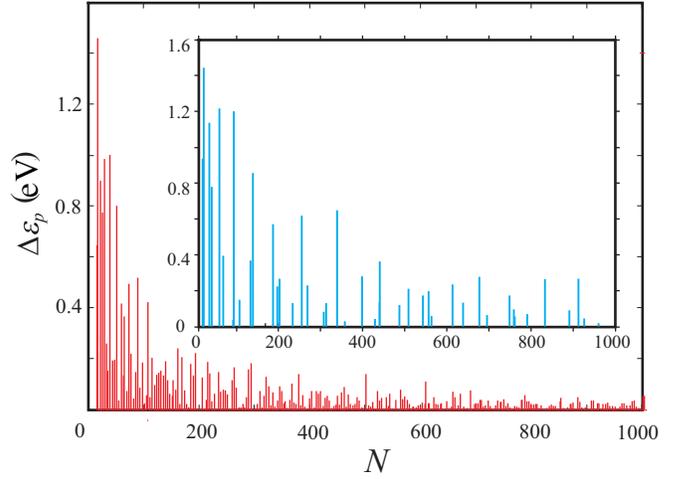}
\caption {Inset shows the specific difference between energies of
lowest unoccupied electron state $\varepsilon^{\rm LU}$ and highest
occupied one $\varepsilon^{\rm HO}$ in neutral discs (red) and
spheres (blue) Au$_{N_{0}}$ at $T=0$.} \label{Fig1}
\end {figure}

The characteristic Coulomb energy of charging is $\widetilde{E}_{\rm
C} =e^{2}/C$, where $C$ is self-capacitance of single granule in a
vacuum (in the case of a disc, the capacitance can be estimated as
for the oblate spheroids of equal volume). The calculations of Ref.
\cite{Nano-2006} demonstrated that these $C$ are too small for the
width of the current gap to be explained. The most obvious example
is the case of a disc, since almost half of the disc surface
contacts to the dielectric film. Therefore, for these granules we
change $C \Rightarrow (1 +\epsilon)C/2$. Then, for discs  and
spheres we have $\widetilde{E}_{\rm C} \simeq \{1.60,\, 0.21\}$ and
$\{1.82,\, 1.06\}$\,eV, respectively. We note that the value of the
capacitance is sensitive to the shape of the granule surface, and
even small deviation from the spherical shape can change
significantly the capacitance.

The consequence of the phonon spectrum deformation of granules is
the weakening of the electron-phonon interaction within them:
$v_{\rm F}/R \gg \omega _{\rm D}$, where $v_{\rm F}$ is the electron
velocity at the Fermi surface in the bulk, and $\omega _{\rm D}$ is
the Debye frequency. This interaction can be so suppressed that the
electron-electron interaction becomes the main mechanism for the
dissipation of the energy, which is injected to the particle. This
additional energy results in the overheating of the electron
subsystem, which is described by the Fermi statistics with some
effective (enhanced) temperature $T^{\rm g}_{\rm eff}$, and the
temperature of the ion subsystem only slightly changes
\cite{Shklov}. With the increase of the bias voltage $V$, the number
of electrons, relaxing in the granule, increases significantly.
\begin{figure}
\centering
\includegraphics [width =.45\textwidth] {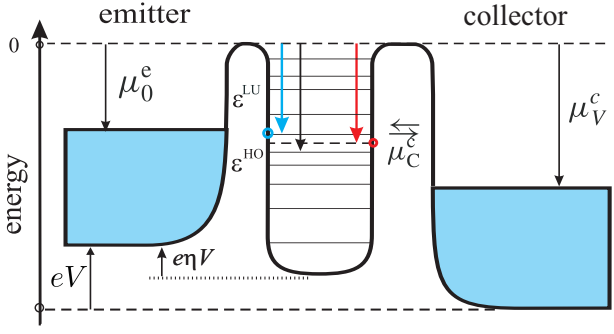}
\caption{Energy profile of structure for $V>0$. $\mu_{\rm C}^{\rm
c}$ is the electron chemical potential of charged granule in
external electric field \cite{Nano-2006}.} \label{Fig2}
\end {figure}

Among them are all the electrons with energies in the interval
$e\eta V$ (see Figure 2) below the Fermi level of the granule ($\eta
V$ is the fraction of the bias voltage on the granule), since the
``flow'' of tunneling electrons increases from below lying levels,
thereby, involving large number of conductivity electrons to the
relaxation process. At the same time, channels of losses appear,
which are related to the generation of holes on the occupied levels
and their subsequent recombination. The granule does not fragmentize
at the significant overheating of the electron subsystem, because
the $I-V$ curves are reproduced at the cyclic changes of the bias
voltage \cite{Wang,Ohgi-2004}.

The estimate of the energy, which is pumped by the conductivity
electrons to the granules of discontinuous films, is given in Ref.
\cite{Tom} ($\sim 0.2,\: 0.3$\,eV). This means that the experiments
\cite{Wang,Ohgi-2004} correspond to the Coulomb blockade regime in
the region of current gap at the whole diapason of $R$ and
reasonable values of $T^{\rm g}_{\rm eff}$. Also, the quantum ladder
can be smeared out by the thermal fluctuations,
$$
\widetilde{E}_{\rm C} > \Delta \varepsilon_{\rm F}\gtrsim k_{\rm
B}T^{\rm e,c,g},
$$
where $\Delta \varepsilon_{\rm F}$ is the difference between
discrete levels in the vicinity of the granule Fermi level, and
$\Delta \varepsilon_{\rm F}=\Delta\varepsilon_{p}$ for magic
clusters at $T=0$.

According to the simple model of Ref. \cite{Nano-2006}, we represent
the emitter and the collector as the electron reservoirs with
continuum spectrums, which are occupied in accordance with the
Fermi-Dirac distribution with chemical electron potential
$\mu_{0}^{\rm e,c}<0$ and temperatures $T^{\rm e,c}$ equal to the
thermostat one. In all cases the energy is counted off from the
vacuum level. The electron chemical potential $\mu^{\rm g}$ of a
granule in a quantum case can be defined by the normalization
condition at a given temperature $T^{\rm g}_{\rm eff}$:
\begin {eqnarray}
\sum_{p=1}^{\infty}\left\{1+\exp[(\varepsilon_{p} - \mu^{\rm
g})/k_{\rm B}T^{\rm g}_{\rm eff}]\right\}^{-1}=N_{0}, \label
{set-1a}
\end {eqnarray}
The summation in (\ref{set-1a}) is performed for all one-particle
states, $N_{0}$ is the average number of conduction electrons in a
granule. The spectrum of states is calculated in advance and,
therefore, the chemical potential of neutral Au$_{N_{0}}$ granules
and its temperature dependence can be found from equation
(\ref{set-1a}).
\begin{figure}
\centering
\includegraphics [width =.49\textwidth]{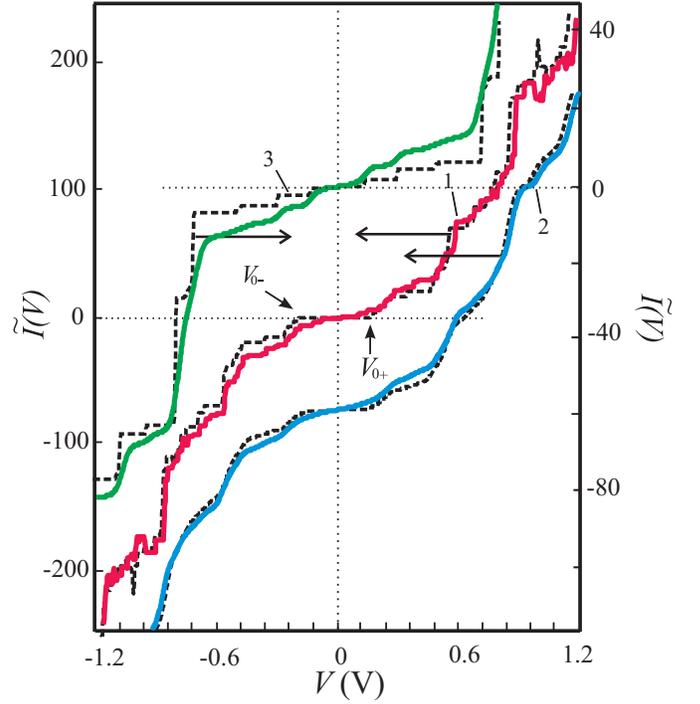}
\caption {Calculated $I-V$ curves of structure based on magic
clusters: disc Au$_{230}$ and sphere Au$_{256}$. 1 -- Au$_{230}$:
dotted curve -- $T^{\rm e,c,g} =5$\,K, solid (red) -- $T^{\rm e,c}
=5$ and $T^{\rm g}_{\rm eff} =2000$\,K. 2 -- Au$_{230}$: dotted
curve -- $T^{\rm e,c,g} =300$\,K, solid (blue) -- $T^{\rm e,c} =300$
and $T^{\rm g}_{\rm eff} =2000$\,K. 3 -- Au$_{256}$: dotted curve --
$T^{\rm e,c} =30$\,K and $T^{\rm g}_{\rm eff} =2000$\,K, solid
(green) -- $T^{\rm e,c}_{\rm eff} =300$ and $T^{\rm g}_{\rm eff}
=2000$\,K.} \label{Fig3}
\end {figure}

The current flowing through a metallic quantum granule (with
restriction on its Coulomb instability \cite{Nano-2006}), is
determined as $I^{\rm e} =I^{\rm c}$ or
\begin{eqnarray}
-e\sum_{n_{\rm min}}^{n_{\rm max}} P_{
n}\left(\overrightarrow{\omega _{ n}^{\rm  e}}-\overleftarrow{\omega
_{ n}^{\rm  e}}\right)= - e\sum_{n_{\rm min}}^{n_{\rm max}} P_{
n}\left(\overrightarrow{\omega _{ n}^{\rm c}}-\overleftarrow{\omega
_{ n}^{\rm  c}}\right), \label{set-60}
\end{eqnarray}
where $P_{ n}$ (the probability of the finding ``in average'' of $n$
surplus  electrons at the granule) is defined by the master equation
in the stationary limit. In fact, the reduced current is calculated
$\widetilde{I} \equiv I/(eP_{0}\Gamma^{\rm e})$, where $\Gamma^{\rm
e,c}$ are the tunnel rates. In order to find values of $P_{n\neq
0}/P_{0}$ we use the recurrent relation.
$\overleftarrow{\overrightarrow{\omega _{ n}}}$ are the partial
electron ``streams'' from the last electrodes to the granule and in
the opposite direction. For the comparison with the results of Ref.
\cite{Wang,Ohgi-2004}, the calculations are done for three
temperatures of the collector and emitter $T^{\rm e} =T^{\rm c}=5,\,
30,\, 300$\,K, and also $T^{\rm g}_{\rm eff} =T^{\rm e},\, 2000$\,K.
The values $\Gamma^{\rm c}=\Gamma^{\rm e}=1$ and $\eta=1/2$ are used
for all cases.

Figure 3 shows calculated $I-V$ curves for disc of radius $R=2$\,nm
(magic cluster Au$_{230}$) and sphere of radius $R=1$\,nm (magic
cluster Au$_{256}$). For low temperatures ($k_{\rm B}T^{\rm g}_{\rm
eff} \ll \Delta \varepsilon_{\rm F}$), the current gap width $\Delta
V_{g} =|V_{0-}|+V_{0+}$ is determined analytically by the
conductance gap boundaries $V_{0-}$ and $V_{0+}$. For example,
$V_{0+}$ is defined from the condition of absence of collector
current of the direct $I-V$ curve branche ( $V>0$), and finally we
have:
\begin{eqnarray}
\Delta V_{g}=\Big(\frac{1}{2e}\tilde{E}_{\rm C} +
\frac{1}{e}\Delta\varepsilon\Big)\Big[\frac{1}{2-\eta}+\frac{1}{1+\eta}\Big],
\label {gap}
\end{eqnarray}
where $\Delta\varepsilon \equiv \mu^{\rm g}-\varepsilon^{\rm HO}
\geq 0$ at $T=0$ and $\eta$ is fixed as for $V>0$. Calculated values
of $\Delta V_{g}$ are in a good agreement with the experimental
values based both on spherical and disc-shape clusters.

The calculation of the $I-V$ curves and current gap can be done only
numerically at $k_{\rm B}T^{\rm g}_{\rm eff} \geq \Delta
\varepsilon_{\rm F}$, when the larger part of the spectrum, compared
to $\Delta \varepsilon_{\rm F}$, is responsible for the charge
transfer. Our calculations show an evident dependence of $I-V$
curves flatness on the electron subsystem temperature.

However, it order to obtain an agreement with observed $I-V$ curves
it is necessary to suggest that electrons in the emitter and
collector are also heated up to some effective temperature, which is
higher than the thermostat one. It is possible, because electrons
(the current $I=1$\,pA is provided by $I/e\sim 10^{6}$ number of
electrons per second) relax in generally on the free path length in
the last electrodes.

For the illustration, we present our result at Figure 3 for the
sphere at $T^{\rm e,c}_{\rm eff}=300$\,K and $T^{\rm g}_{\rm
eff}=2000$\,K. Only by such a way, we can explain the flattening of
the $I-V$ curves for the metallic cluster structures at low
thermostat temperatures. With the increase of the bias voltage, the
current flow is accompanied by the increase of the electron gas
temperature.

In conclusion, we have calculated the $I-V$ characteristics of
structure based on magic clusters: disc Au$_{230}$ and sphere
Au$_{256}$. We have suggested that the overheating of electron
subsystem leads to the disappearance of current gap and significant
flattening of current--voltage curves. Our results are in a good
qualitative agreement with experiment of Ref. \cite{Wang,Ohgi-2004}.

\acknowledgments{We are grateful to W. V. Pogosov for reading the
manuscript. This work was supported by the Ministry of Education and
Science of Ukraine (Programme "Nanostructures") and Samsung
Corporation.}

\end{document}